# Newly Discovered Eclipsing Binary
# 2MASS J18024395+4003309 (VSX J180243.9+400331):
# Two-Color Photometry vs Phenomenological Modeling

Ivan L. Andronov[1], Yonggi Kim[2], Young-Hee Kim[2], Joh-Na Yoon[2],
Lidia L. Chinarova[3], Mariia G. Tkachenko[1]

[1] Department "High and Applied Mathematics", Odessa National Maritime University, Odessa, Ukraine

[2] Astronomical Observatory, Chungbuk National University, Korea

[3] Astronomical Observatory, I.I.Mechnikov Odessa National University, Odessa, Ukraine

**Abstract.** We report on analysis of the two-color VR CCD observations of the newly discovered variable 2MASS J18024395+4003309 = VSX J180243.9 +400331 obtained using the 1-m telescope of the Mt. Lemmon Observatory (LOAO) in the field of the intermediate polar V1323 Her. The extended version of this conference talk we published in 2015JASS...32..127A. The variability was reported in 2012OAP....25..150A, and the object was monitored. The two-color observations covered all phase interval. The object is classified as an Algol-type variable with tidally distorted components, and shows an asymmetry of the maxima (the O'Connell effect). For phenomenological modeling, we used the trigonometric polynomial approximation of statistically optimal degree, and a recent method "NAV" (New Algol Variable) using local specific shapes for the eclipse. Methodological aspects are described, especially for the case of few color observations. Estimates of the physical parameters based on analysis of phenomenological parameters, are presented.

## Introduction

For the phenomenological modeling of eclipsing binary stars, Andronov (2012a) proposed the "New Algol Variable" algorithm. It was also presented in the "Czestochowski Kalendarz Astronomizny – 2013" by Andronov (2012b). The method was improved for the case of unknown period and initial epoch, which may be also independent variables, by Andronov et al (2015a). The applications to 6 more stars was presented by Tkachenko et al. (2015).

The variability of the object 2MASS J18024395 +4003309, which is located in the field of the intermediate polar V1323 Her by Breus (2012). This object got a name VSX J180243.9+400331 (hereafter shorter VSX1802) and an object identifier 282837. Unfortunately, no GCVS name is given to this object yet.

Due to a relatively large angular distance (14') from V1323 Her, VSX1802 is seen in the field of V1323 Her only at the CCD images with a focal reducer. This was the case for one night reported by Andronov et al. (2012), who noted a sharp profile of the minimum. Additional observations from the Catalina survey (Drake et al. 2009) allowed them to determine photometric elements $T_0$=2456074.4904, $P$=0.3348837±0.0000002$^d$.

### Observations and Periodogram Analysis

Due to the installation of the focal reducer at the 1-m telescope of the Mt. Lemmon Astronomical Observatory, Korea (LOAO), the field of CCD observations increased and allowed studies of objects in the field, in an addition to the main targets (in this case, the intermediate polar V1343 Her). Totally we obtained 196 observations in V (range 16.51$^m$ – 17.51$^m$) and 242 observations in R (range 15.88$^m$ – 16.77$^m$). The total duration of observations was 45.5 hours during 11 nights in R and 8 nights in the alternatively changing filters VR. To improve accuracy of the calibration, we have used the method of "artificial comparison star" (Andronov and Baklanov 2004, Kim et al. 2004).

For the periodogram analysis, we have used the trigonometric polynomial fit of a degree $s$:

$$x_C(t) = C_1 + \sum_{j=1}^{s}(C_{2j} \cdot \cos 2\pi f t + C_{2j+1} \cdot \sin 2\pi f t), \qquad (1)$$

where coefficients $C_\alpha, \alpha = 1..m, m = 1 + 2s$ are computed using the least squares method and the test function $S(f)$ was used (cf. Andronov 1994, 2003), and $f$ is frequency (in cycles/day) . For eclipsing binaries, the peaks at odd multipliers of the orbital frequency are typically much smaller than that for even multipliers. For the filters V and R, the peaks occur at similar values, which are close within error estimates to the value $f=2.986111\pm0.000002$ cycles/day (Andronov et al. 2012), so we adopted their ephemeris:

$$\text{Min I. BJD} = 2456074.4904 + 0.3348837 \cdot E \qquad (2).$$

We confirm cycle numbering and mention that the true frequency differs by a value of 0.5/year from the estimate based on 3 minima by Parimucha et al. (2012).

To determine statistically optimal degree $s$ of approximation (1), we have used different criteria (see Andronov 1994, 2003 for detailed discussion). According to the classical Fischer's criterion, we got $s=14$ for the filter V and $s=8$ for R, with a false alarm probability of $FAP=10^{-3}$. Criterion of the best r.m.s. accuracy corresponds to $s=6$, i.e. much smaller value. The Gibbs effect causes apparent waves at the out-of-eclipse part of the light curve, which are not physically justified. In the trigonometric polynomial approximation with $s=6$, the frequency $f$ is a free parameter determined using differential corrections (Andronov, 1994). The best fit estimate of periods is $P= 0.3348842^d \pm 0.0000005^d$ (filter V), $0.3348845^d \pm 0.0000004^d$ (R). The difference from the value in (1) is not statistically significant, but a weighted mean is $0.3348844^d \pm 0.0000003$. The mean moments of the primary minima are $2456238.9186 \pm 0.0005^d$ (V), $2456320.9663 \pm 0.0007^d$ (R). They correspond to different cycles, as the nights in V and R are generally different, and Andronov (1994) argued that the best accuracy estimates are for moments, which are closest to the sample mean of time. For such approach, there is no necessity that the moments should be in the interval of observations in some particular night.

### Phenomenological vs. Physical Modeling

For physical modeling, usually the method of Wilson and Devinney (1971) is used, which was realized in some computer programs (e.g. Zola et al. 1997, 2010). However, for the majority of stars, including the studied one, we do not have spectroscopic observations needed to determine the mass ratio and temperatures. So one may try to make estimates using phenomenological modeling.

Using the NAV algorithm, we have determined approximations for a grid of values of the eclipse width $\Delta\phi = C_8$ with a determination of the parameters describing the shape of eclipse

$$H(z,\beta) = \begin{cases} (1 - |z|^\beta)^{3/2}, & \text{if } |z| < 1 \\ 0 & \text{else} \end{cases}$$

The dimensionless variable $z$ is related to phase $\phi$ as

$$z = \frac{(\phi-\phi_0) - \text{int}(\phi-\phi_0+0.5)}{\Delta\phi}.$$

For $\beta=0$, the shape is narrow and is physically unrealistic, for $\beta=1$, the shape at the center of eclipse is triangular, for $\beta=2$ is parabolic, and for $\beta \to \infty$ tends to a rectangle.

Review on the methods for determination of the extrema was presented by Andronov (2005). The complete theory of statistical properties of the smoothing function in a case of arbitrary functions and additional (wavelet-like) time- and scale-dependent weight functions was presented by Andronov (1997). Other shapes for phenomenological modeling of eclipses based on modifications of the Gaussian are discussed by Mikulášek et al. (2012) and Mikulášek (2015).

To get a single value of $C_8$ for two filters, we made a scaled sum of the test functions

$$\Phi(C_8) = \frac{\Phi_V(C_8)}{\Phi_{V,min}} + \frac{\Phi_R(C_8)}{\Phi_{R,min}}$$

and determined the filter half-width $C_8=0.1177$ corresponding to a minimum.

The classical phenomenological parameters listed in the GCVS (Samus et al. 2014) were determined (Andronov et al. 2015a): Max I = $16.567^m \pm 0.006^m$, Max II = $16.592^m \pm 0.006^m$, Min I = $17.493^m \pm 0.014^m$, Min II = $17.281^m \pm 0.008^m$, Min I-Max I= $0.926^m$, Min II-Max I= $0.714^m$ (filter V).

These phenomenological parameters may be used for estimates of the physical parameters. For this, one may use the "simplified" model of spherical components without effects of ellipticity, reflection and limb darkening (Kopal 1959, Shulberg 1970, Tsessevich 1971, Malkov et al. 2007, Andronov and Tkachenko 2013). If using the phenomenological parameters mentioned above, the corresponding point at the "depth – depth" diagram (Fig.1 in Malkov et al. 2007) lies close to the line "R" of equal radii, but slightly outside the allowed region. This is obviously explained because more correct is to use the "NAV" parameters $C_6$ and $C_7$ instead of the classical Min I-Max I and Min II-Max I.

The sharp shape of the eclipses also argues for similar characteristics of the two stars. Assuming that the components are Main Sequence stars, it is possible to use statistical "Mass-Radius-Luminosity" relations (e.g. Allen 1973, Cox 1980) and to apply them for the analysis along with $C_6$ and $C_7$.

In our case of two-color photometry, one may check the solutions for each color. Andronov et al (2015a) discussed the procedure in detail. Here we would like to note that the results are self consistent, with only few per cent of uncertainty, mainly due to different estimates of the statistical dependencies. Using the oversimplified form of the MS mass-radius relation $R = \frac{R_\odot \cdot M}{M_\odot}$ (Faulkner 1971) and, combining with the third Kepler's law, we obtain for the inclination $i = 90°$, $M_1 = 0.745$, $M_2 = 0.854$, $M = M_1 + M_2 = 1.599$, the orbital separation $a=1.65\cdot10^9$m$=2.37R_\odot$ and relative radii $r_1=R_1/a=0.314$ and $r_2=R_2/a=0.360$.

Generally, these are minimal estimates of the radiuses and masses, as for the inclination $i$, differing from 90°, we'll get larger estimated values. In the assumption of one total eclipse, one may write cos $i<|r2-r1|=0.046$, thus 87.4°≤i≤ 90°, 0.999≤sin i≤1, so for VSX 1802 the effects of inclination onto estimates of radii and masses are negligible. According to Cox (2000), the spectral classes of the stars are G8 and K2 for the mass estimates mentioned above.

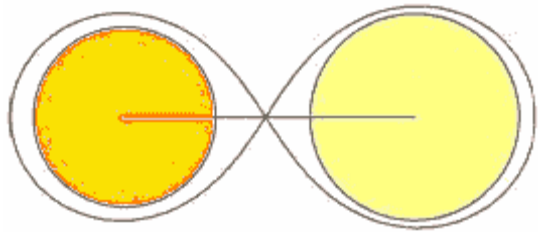

**Figure 1**. The model of the system VSX 1802: The Roche lobes, the line of centers and circles corresponding to estimated radii of the stars in a spherically-symmetric approximation.

*Acknowledgements.* We thank Dr. Bogdan Wszołek and LOC for the excellent hospitality during the conference "Astrophysica Nova" and Prof. Stanisław Zoła for fruitful discussions. This work is also a part of the "Inter-Longitude Astronomy" campaign (Andronov et al. 2010) and a project "Ukrainian Virtual Observatory (Vavilova et al. 2012).